\documentclass[aps,pra,amsmath,amssymb,notitlepage,reprint,10pt,longbibliography,superscriptaddress]{revtex4-2}
\usepackage{amsmath,epsfig,amssymb}
\usepackage{bbm}
\usepackage{times}
\usepackage{color}
\usepackage{amsthm}
\usepackage{xcolor}
\usepackage[caption=false]{subfig}
\usepackage{multirow}
\usepackage[normalem]{ulem}
\usepackage{amsfonts,amssymb,dsfont,bm}
\usepackage{graphicx}
\usepackage{soul}
\usepackage{hyperref}
\hypersetup{
    colorlinks=true,       
    linkcolor=red,          
  citecolor=magenta,        
    filecolor=magenta,      
    urlcolor=cyan,           
    runcolor=cyan
}
  
\usepackage[acronym,shortcuts]{glossaries}

\newcommand{\todo}[1]{}
\renewcommand{\todo}[1]{{\color{red} TODO: {#1}}}
\newcommand{\note}[1]{}
\renewcommand{\note}[1]{{\color{red} NOTE: {#1}}}
\newcommand{\question}[1]{}
\renewcommand{\question}[1]{{\color{red} QUESTION: {#1}}}

\newcommand\sinc{\operatorname{sinc}}

\newacronym{ARMA}{ARMA}{autoregressive moving-average}
\newacronym{SchWARMA}{SchWARMA}{Schr\"{o}dinger Wave ARMA}
\newacronym{MUSIC}{MUSIC}{multiple signal classification}
\newacronym{QNS}{QNS}{qubit noise spectroscopy}
\newacronym{PSD}{PSD}{power spectral density}
\newacronym{NNLS}{NNLS}{nonnegative least squares}
\newacronym{YW}{YW}{Yule-Walker}
\newacronym{FTTPS}{FTTPS}{fixed total time pulse sequences}
\newacronym{AIC}{AIC}{Akaike information criterion}
\newacronym{BIC}{BIC}{Bayesian information criterion}
\newacronym{MSE}{MSE}{mean-squared error}

\begin{document} 

\glsdisablehyper

%
\title{Model-Based Qubit Noise Spectroscopy}
\author{Kevin Schultz}
\affiliation{Johns Hopkins University Applied Physics Laboratory, Laurel, MD 20723, USA}

\author{Christopher A.~Watson}
\affiliation{Johns Hopkins University Applied Physics Laboratory, Laurel, MD 20723, USA}

\author{Andrew J.~Murphy}
\affiliation{Johns Hopkins University Applied Physics Laboratory, Laurel, MD 20723, USA}

\author{Timothy M.~Sweeney}
\affiliation{Johns Hopkins University Applied Physics Laboratory, Laurel, MD 20723, USA}

\author{Gregory Quiroz}
\affiliation{Johns Hopkins University Applied Physics Laboratory, Laurel, MD 20723, USA}
\affiliation{William H. Miller III Department of Physics $\&$ Astronomy, The Johns Hopkins University, Baltimore, MD 21218, USA}

\begin{abstract}

Qubit noise spectroscopy (QNS) is a valuable tool for both the characterization of a qubit's environment and as a precursor to more effective qubit control to improve qubit fidelities. Existing approaches to QNS are what the classical spectrum estimation literature would call ``non-parametric'' approaches, in that a series of probe sequences are used to estimate noise power at a set of points or bands. In contrast, model-based approaches to spectrum estimation assume additional structure in the form of the spectrum and leverage this for improved statistical accuracy or other capabilities, such as superresolution.
Here, we derive model-based QNS approaches using inspiration from classical signal processing, primarily though the recently developed Schr\"{o}dinger wave autoregressive moving-average (SchWARMA) formalism for modeling correlated noise. We show, through both simulation and experimental data, how these model-based QNS approaches maintain the statistical and computational benefits of their classical counterparts, resulting in powerful new estimation approaches.
Beyond the direct application of these approaches to QNS and quantum sensing, we anticipate that the flexibility of the underlying models will find utility in adaptive feedback control for quantum systems, in analogy with their role in classical adaptive signal processing and control.

\end{abstract}

\maketitle

\section{Introduction}
Fault-tolerant error-corrected quantum computation offers the potential for more efficient computation in a number of problem domains, such as factoring of numbers using Shor's algorithm \cite{shor1994algorithms,chuang1995quantum,shor1999polynomial}, 
solving linear systems \cite{harrow2009quantum,clader2013preconditioned,pan2014experimental,wossnig2018quantum},
quantum simulation and chemistry \cite{cirac2012goals,jones2012faster,houck2012chip,georgescu2013quantum,o2016scalable},  and other areas \cite{shor2002introduction,childs2010quantum,venegas2012quantum,schuld2015introduction,montanaro2016quantum,biamonte2017quantum}.  Key to this fault tolerance are high-fidelity physical gates whose error-rates are below some threshold level, such that repeated layers of error correction can arbitrarily reduce logical error rates. Considerable research has been performed on reducing error sources present in a device (through, e.g., novel materials, process improvements, and circuit design \cite{chang2013improved,oliver2013materials,venturelli2018compiling,kusyk2021survey,mergenthaler2021effects}), as well as mitigating the impact of error sources using quantum control \cite{1998BanPhotonEchoDD,1998ViolaDynamicalErrorSupp,1999ViolaDDOpenSystems,PhysRevLett.113.250501,soare2014experimental,paz2016dynamical,PhysRevResearch.2.013156}. 

One particularly impactful application of quantum control is the reduction of semiclassical, time-correlated noise. This type of noise is widely present in physical systems \cite{bylander2011:fnoise, yan2013:fnoise,burnett2014:fnoise,muller2015:fnoise,burnett2019:fnoise,basset2014:fnoise,chan2018:qns,struck2020:fnoise}. In order to maximize the efficacy of quantum control, it is generally necessary to first characterize the noise environment, a process called \ac{QNS} \cite{alvarez2011:qns,PhysRevLett.116.150503,szakowski2017:qns,pazsilva2017:qns,norris2018optimally,PhysRevA.100.042334,sung2019non,PRXQuantum.1.010305}. These approaches can be interpreted through the lens of the filter function formalism \cite{cywinski2008:fff,PhysRevLett.113.250501}, which relates a given probe sequence to its filter function in the frequency domain, and the corresponding observable to the overlap of the filter function with the noise power spectrum. Many different probe sequences are then fed through the system, and the expectation values of the measured observables are jointly analyzed (typically through a form of least squares regression analysis) to estimate the power spectrum of the noise.  
 
In the context of \textit{classic} spectrum estimation, these \ac{QNS} approaches closely resemble ``nonparametric'' spectrum estimation. These estimators can be interpreted as taking a collection of filters, applying them to the signal of interest, and analyzing the collective outputs as inner products (i.e., overlaps) to estimate the power spectrum \cite{mullis1991quadratic}. Nonparametric approaches to spectrum estimation include the periodogram (squared magnitude of the (fast) Fourier transform of the signal), windowed and averaged periodograms (such as Bartlett's \cite{bartlett1955efficiency} and Welch's \cite{welch1967use} methods), and the multitaper method \cite{thomson1982spectrum} (c.f., the qubit control noise estimation considered in \cite{norris2018optimally}). 

In contrast, ``parametric'' or model-based approaches to noise spectroscopy \cite{kay1981spectrum,stoica2005spectral} assume a functional form (i.e., a model) and attempt to estimate the parameters rather than the spectrum directly. Model-based approaches offer statistical benefits over nonparametric approaches, such as reduced error variance (fewer parameters to estimate with the same data) and flexible approaches to model selection to reduce overfitting. Beyond these statistical benefits, we note that these model-based approaches produce estimates defined at all frequency points, as compared to the finite band resolution in nonparametric approaches, offering the potential for so-called superresolution.

Here, we adapt some common model-based approaches from classical noise spectroscopy, specifically \ac{ARMA} models and line spectra, to the \ac{QNS} problem. In addition to direct adaptation of classical algorithms, such as \ac{YW} \cite{udny1927method,walker1931periodicity,box2015time} and \ac{MUSIC} \cite{schmidt1986multiple}, we show how the recently introduced \ac{SchWARMA} model \cite{schultz2021schwarma,murphy2021universal} can be used to directly fit model parameters and how model-selection rules from classical spectrum estimation can be used to reduce overfitting and improve estimation accuracy. Beyond \ac{QNS}, we note that these approaches may find applications in related problems in quantum sensing and control, c.f., the line spectrum signal model of Ref.~\cite{poggiali2018optimal}, bandlimited signal model of Ref.~\cite{titum2021optimal}, or control noise model of \cite{trout2022provably}.

In this paper, Sec.~\ref{sec:qns} reviews \ac{QNS} in the context of single-qubit dephasing noise, although we note that the approaches here should extend to arbitrary noise scenarios where filter-function based approximations remain valid, such as control noise \cite{norris2018optimally}, multi-qubit noise \cite{pazsilva2017:qns,PRXQuantum.1.010305}, multi-axis noise \cite{PhysRevA.100.042334}, non-Gaussian noise \cite{PhysRevLett.116.150503,sung2019non}, and quantum noise \cite{pazsilva2017:qns}. In Sec.~\ref{sec:mbse}, we formally introduce the most common models used in classical spectrum estimation and discuss how to adapt common fitting algorithms to the \ac{QNS} case, also introducing a flexible new approach based on the \ac{SchWARMA} model. Next, in Sec.~\ref{sec:se}, we demonstrate the efficacy of model-based \ac{QNS} using \ac{SchWARMA} via simulation in white noise, bandlimited noise, and complex $1/f^\alpha$-like spectra. We then consider experimental bandlimited noise injection data taken from Ref.~\cite{murphy2021universal} to further validate the \ac{SchWARMA}-fitting approach. We then pivot in Sec.~\ref{sec:superres} to consider a different application of model-based \ac{QNS}, where we show how our techniques are able to finely resolve spectral features using both simulated and experimental data.   We conclude with discussion and potential future directions. 

\section{Qubit Noise Spectroscopy}\label{sec:qns}
In this work, we are concerned with the estimation of a dephasing noise spectrum of a qubit, i.e., a qubit undergoing semi-classical noisy Hamiltonian dynamics governed by
\begin{equation}\label{eq:depham}
    H(t) = \eta(t)\sigma_z/2+\Omega(t)\sigma_x/2
\end{equation}
where $\eta$ is a zero-mean, wide-sense stationary Gaussian process and $\Omega$ is our single axis control signal.  In this formulation, the role of $\Omega(t)$ is to define a series of noise probe sequences that can estimate the power spectrum $S_\eta(\omega)$ of $\eta$, i.e., the Fourier transform of the autocovariance $E[\eta(t_1),\eta(t_2)]=E[\eta(|t_1-t_2|)\eta(0)]\triangleq r_\eta(t_1-t_2)$.

The relationship between the control and the noise spectrum of the model in Eq.~\eqref{eq:depham} can be interpreted through the filter function formalism \cite{cywinski2008:fff,PhysRevLett.113.250501}. In the restricted setting of Eq.~\eqref{eq:depham}, we can associate with the control sequence $\Omega(t)$ a frequency domain representation $F_\Omega(\omega)$ of the control's impact on the dephasing noise and approximate the survival probability of a qubit prepared and measured in the $|+\rangle$ state by 
\begin{equation}\label{eq:surv}
  p\approx \frac{1}{2}+\frac{1}{2}\exp\left(-\int\,d\omega S_\eta(\omega)F_\Omega(\omega)\right)\,.
\end{equation}
The quantity $\chi=\int\,d\omega S_\eta(\omega)F_{\Omega}(\omega)$ is often referred to as the \textit{overlap integral} in the literature.

Eq.~\eqref{eq:surv} suggests a method for estimating the power spectrum $S_\eta$ by using different control sequences $\Omega_k$ to measure survival probabilities $p_k$ and then ``inverting'' the relationship in \eqref{eq:surv}.  Since the left hand side of Eq.~\eqref{eq:surv} is discrete, but $S_\eta(\omega)$ is continuous in $\omega$, it is common to restrict the estimation to discrete frequency points $\omega_j$.  Assuming $K$ total sequences and $J$ total frequency points to be estimated, this sets up a system of equations
\begin{equation}
\chi_k \approx \sum_{j=1}^J F_{\Omega_k}(w_j)S_\eta(\omega_j)
\end{equation}
where  $\chi_k=\log\left(2\left(p_k-\frac{1}{2}\right)\right)$ is measured experimentally. This can be expressed in matrix form as
\begin{equation}
  \bm{\chi}\approx \bm{F}\bm{S}\,,
\end{equation}
which can be solved using some form of linear regression to recover an estimate $\hat{\bm{S}}$ of the power spectrum.  Assuming $J\geq K$ and $\bm{F}$ is reasonably well conditioned, this can be solved using \ac{NNLS} to recover an estimate $\hat{\bm{S}}$. Here, we consider the \ac{FTTPS} used in \cite{schultz2021schwarma,murphy2021universal}, which are $X(\pi)$-pulse sequences that result in spectral concentration at fixed intervals, thus $\omega_j=j\omega_*$ for some $\omega_*$ determined by the gate-duration and overall probe sequence length.

\section{Model-Based Spectrum Estimation}\label{sec:mbse}
In classical signal processing, spectrum estimation of a time series is a well-studied problem \cite{bartlett1955efficiency,welch1967use,kay1981spectrum,thomson1982spectrum,mullis1991quadratic,stoica2005spectral}. Fourier-analytic approaches to spectrum analysis date back at least as far as the late 19th century with the work of Schuster \cite{schuster1898investigation}, who appears to have originally coined the term \textit{periodogram} \cite{schuster1899periodgram}.  Unfortunately, a ``raw'' periodogram (that is, the square of the amplitudes of a discrete-time Fourier transform of a time series) does not reduce the variance of a spectrum estimate as the number of samples increases.  Instead, techniques based on refining the periodogram, such as Blackman-Tukey's, Bartlett's, Welch's, Daniell's, and Thomson's methods \cite[Ch.~2]{stoica2005spectral}, use some additional form of averaging and windowing (i.e., filtering) to produce statistically well-behaved estimates in a coarser set of frequency bands. All of these approaches are considered nonparametric, as they place no additional assumptions on the signal model (other than wide sense stationarity and Gaussianity) and can be directly interpreted as estimating the power in a set of frequency bands using a set of bandpass filters \cite{mullis1991quadratic}.

In contrast, methods that assume the signal satisfies a generating model with known functional form and attempt to estimate parameters of that model are referred to as parametric or model-based approaches \cite{kay1981spectrum,stoica2005spectral}. 
In the context of the filter-based viewpoint of nonparametric estimators \cite{mullis1991quadratic}, parametric estimators do not have to consider sidelobe or leakage effects that can impact the frequency concentration in the passband. These model-based approaches implicitly incorporate this model as additional information and can potentially offer statistical benefits over nonparametric estimators when the number of model parameters to be estimated is fewer than the number of nonparametric frequency bands required to achieve a desired spectral resolution.  Additionally, model-based approaches offer natural options for model-selection \cite{stoica2004model} that reduce chances of overfitting.

\subsection{Classical Methods for Model-Based Spectrum Estimation}
There are a number of classical methods for spectrum estimation that fall under the designation of model-based estimates.  Perhaps the most common approach to model-based spectrum estimation is to fit the time series in question to an \ac{ARMA} model, which models a classical discrete-time series $y[k]$ via the linear combination of $p$ prior values $y[k-i]$ and $q+1$ independent Gaussian inputs from the time series $x[k-j]$ via
\begin{equation}\label{eq:ARMA}
   y[k] = \underbrace{\sum_{i=1}^p a_{i}
    y[k-i]}_{AR}+\underbrace{\sum_{j=0}^{q} b_{j}x[k-j]}_{MA}\,,
\end{equation}
which is denoted an $ARMA(p, q)$ model (for $p,q>0$).  When the model has $q=0$, it is known as an $AR(p)$ (autoregressive of order $p$) model, and when $p=0$ it is known as an $MA(q)$ (moving-average model of order $q$).  An \ac{ARMA} model results in a (discrete-time) power spectrum $S_y(\omega)$ for $y$ of
\begin{equation}\label{eq:ARMAPSD}
    S_y(\omega)=\frac{\left|\sum_{k=0}^q  b_k \exp(-ik\omega)\right|^2}{\left|1+\sum_{k=1}^pa_k\exp(-ik\omega)\right|^2}\,.
\end{equation}
In other words, the \ac{ARMA} approach assumes a rational polynomial form of the power spectrum, but this restriction is known to be dense in the space of valid power spectra \cite{stoica2005spectral} (periodic in $2\pi$, as for discrete-time signals).  Furthermore, this density property holds for pure AR and MA models as well.

Another model common in the classical literature is to assume that the signal has a \textit{line spectrum}, that is, the signal $y(t)$ is a sum of sinusoids (the signal $x(t)$) in the presence of additive noise $e(t)$:
\begin{equation}
  y(t)=\underbrace{\sum_{k=1}^n A_k\exp(i(\zeta_kt+\phi_k))}_{x(t)}+e(t)\,,
\end{equation}
where $A_k$, $\zeta_k$, and $\phi_k$ are the unknown (to be inferred) amplitude, frequency, and phase, respectively, of the $k$th sinusoid.  
We note that for the purposes of power spectrum estimation, the phases are irrelevant.
Generally, $n$ is assumed to be specified or otherwise selected based on the data (as with the \ac{ARMA} model orders $p$ and $q$, above), and $e(t)$ is generally assumed to be white Gaussian noise.  

\subsection{Adaptation to QNS}
Unlike the classical case, where one would have access to the complete time series, in the \ac{QNS} setting we have access to only some collection of sample average values. Via the filter function formalism, we see that this is approximately the average of the overlap between the control signal's filter ``power'' and the dephasing noise spectrum.  In this sense, conventional \ac{QNS} approaches most closely resemble the quadratic estimator viewpoint of classical filtering/windowing-based methods \cite{mullis1991quadratic}, albeit with constraints on the available forms of the filter to be used. 

To overcome this limitation, we propose two different methods for adapting these classic model-based approaches to the quantum setting.  The first approach is to estimate, using \ac{NNLS} \ac{QNS}, the noise power spectrum $\hat{S}_\eta$, or equivalently via Fourier transform the autocovariance $\hat{r}_\eta$. The spectrum can then be used to produce model-based spectrum estimates via approaches that do not require access to the time series itself, such as \ac{YW} approaches to AR models.
The second option, which applies to general (i.e., AR-, MA-, or ARMA-based) models, is to exploit the differentiability of Eqs.~\eqref{eq:ARMAPSD} and \eqref{eq:surv} using the recently introduced \ac{SchWARMA} model \cite{schultz2021schwarma}.

\subsection{Spectrum Estimation Using $\hat{r}_\eta$}
Using the \ac{NNLS} approach to spectrum estimation, one can produce an estimate $\hat{S}_\eta(\omega_j)$ at the discrete frequencies $\omega_j$.  By applying the inverse discrete-time Fourier transform to $\hat{S}_\eta$, we can obtain an estimate of the discrete-time autocovariance function $\hat{r}_\eta(t_j)$.  Classically, many parametric techniques for discrete-time power spectrum estimation are derived from the autocovariance, as it is a trivial calculation to estimate the variance between a time series and time-lagged versions of itself. In particular, there are ways to estimate the coefficients of an \ac{ARMA} model from $\hat{r}_\eta$ (as opposed to e.g., maximum likelihood methods that rely on samples of the time-series directly). We will briefly review a few of these approaches that we have currently implemented in our software toolbox, \texttt{mezze} \cite{mezze}.

The first approach we will consider is the adaptation of the \ac{YW} equations for estimating $AR(p)$ models to the \ac{QNS} setting.   Classically, the autocovariance $r_y$ of the sequence $y_k$ can be estimated using direct empirical averaging of $\hat{r}_y[k]=\frac{1}{N-j}\sum_{k=0}^{N-j-1}y[k]y[k+j]$ or, more computationally efficiently for large $N$, repeated application of the fast fourier transform. Regardless of the specific method used, it is clear that this is not possible in the QNS context, as we do not have direct access to discrete-time samples of the noise process $\eta$.  Instead, we will use the Fourier transform of the \ac{NNLS} estimate $\hat{S}_\eta$ to produce an estimate $\hat{r}_\eta[k]$.  The \ac{YW} approach for \ac{QNS} then proceeds in the usual fashion by first solving for $a_1,\dots,a_p$ using the Toeplitz autocovariance matrix $\hat{\mathbf{R}}$
\begin{equation}\label{eq:YW}
  \begin{bmatrix}\hat{r}_\eta[1]\\\hat{r}_\eta[2]\\\vdots\\\hat{r}_\eta[p]\end{bmatrix}=
  \underbrace{\begin{bmatrix}\hat{r}_\eta[0]&\hat{r}_\eta[1]&\hat{r}_\eta[2]&\cdots\\
    \hat{r}_\eta[1]&\hat{r}_\eta[2]&\hat{r}_\eta[3]&\cdots\\
    \hat{r}_\eta[2]&\hat{r}_\eta[3]&\hat{r}_\eta[4]&\cdots\\
    \vdots&\vdots&\vdots&\ddots\\
    \hat{r}_\eta[p-1]&\hat{r}_\eta[p-2]&\hat{r}_\eta[p-3]&\cdots\end{bmatrix}}_{\hat{\mathbf{R}}}
  \begin{bmatrix}a_1\\a_2\\a_3\\\vdots\\a_{p}\end{bmatrix}
\end{equation}
in the least squares sense.  Next, the sole remaining unknown in the $AR(p)$ model $b_0$ can be solved by noting that $\hat{r}_\eta[0]=\sum_{k=1}^p a_k\hat{r}_\eta[k]+b_0^2$, and noting that the sign of $b_0$ is irrelevant in the power spectrum in Eq.~\eqref{eq:ARMAPSD}. In addition to this form of \ac{YW}, we will also take advantage of the fact that we will always have $N$ time estimates for $\hat{r}_\eta$ (i.e., in contrast with the classical case, we do not have the option of estimating \textit{only} the first $p$ time lags); we can instead consider the overdetermined set of equations by using the first $p$ columns of $\hat{\mathbf{R}}$ and adapting Eq.~\eqref{eq:YW} accordingly.  A brief discussion on the implications of using an overdetermined \ac{YW} fit can be found in \cite[Sec.~3.7.1]{stoica2005spectral}.

Unfortunately, MA models cannot be cast in a similar linear regression problem.  Instead, we will exploit the fact that an $MA(q)$ model 
\begin{equation}
y[k]=\sum_{j=0}^q b_jx[k-j]
\end{equation}
satisfies $r_y[k] = \sum_{j=k}^qb_jb_{j-k}$, which in particular implies that the autocovariance of $y$ vanishes after $q$ time lags.  Minimization of the function $\sum_{k=0}^q (\hat{r}_\eta[k]-\sum_{j=k}^qb_jb_{j-k})^2$ can be solved using Newton's method or some other form of gradient descent from a suitable initial condition (here $b_0=\hat{r}_\eta[0]+2\sum_{j=1}^q\hat{r}_\eta[j]$, $b_j=0$ for $j>0$) \cite{wilson1969factorization}.  This method converges sufficiently rapidly and accurately for our purposes, but other methods for the design of MA models can be found in the filter design literature \cite{wu1999fir}.

For the fitting of full $ARMA(p,q)$ models, we turn to the cepstrum (log of the power spectrum) recursion techniques of Ref.~\cite{kaderli2000spectral}.  The first step in this approach is to estimate the $AR(p)$ portion of the model using the (overdetermined) \ac{YW} approach as described above.  Defining the cepstrum as the inverse Fourier transform of the log of the power spectrum determined from the \ac{NNLS} estimate, $\hat{c}(\omega_k)=\mathcal{F}^{-1}\{\log(\hat{S}_\eta(\omega_k))\}$, the normalized coefficients $b'_k$ can be estimated recursively by setting $b'_0=1$ and for increasing $k$ (up to $k=q$) via
\begin{equation}\begin{aligned}
b'_k &= \frac{1}{k}\biggl[k\hat{c}_kb'_0-\sum_{m=1}^{k-1}mb'_ma_{k-m}+\sum_{m=1}^kma_mb'_{k-m} \\&+ \sum_{j=1}^{k-1}j\hat{c}_j\sum_{m=0}^{k-j}a_mb'_{k-j-m}\biggr]\,.
\end{aligned}\end{equation}
The above MA coefficients are ``accurate'' up to an overall scale factor $\kappa^2$ which can be determined from the ratio between the \ac{NNLS} estimate and the ARMA power determined by $\{a_k\}$ and $\{b'_k\}$.  We further optimize $\kappa^2$ by finding the minimum mean squared error between the corresponding filter function predictions and the observed outcomes.  We then set the actual MA estimates by $b_j=\kappa b_j'$.

With respect to line spectra, a common approach to spectrum estimation is to fit the spectrum using an $AR(p)$ model \cite{stoica2005spectral}.  Under this model, the complex zeros of the denominator in the rational power spectrum (the so-called ``poles'' of the AR model) will presumably match the frequencies of the line spectrum (i.e., their complex argument corresponds to the normalized frequency) and the power at that frequency determines the amplitude of that sinusoidal component.  In the \ac{QNS} context, this approach can be estimated using the same \ac{YW} adaptation as above.

Beyond the \ac{ARMA} family, there are fundamentally different model-based methods for line spectrum estimation that use estimates of the autocovariance, and here we will consider \ac{MUSIC} \cite{schmidt1986multiple}.  Unlike \ac{ARMA}-based approaches which assume a rational polynomial form of the power spectrum, \ac{MUSIC}  generalizes Pisarenko's method \cite{pisarenko1973retrieval} and instead assumes that the unknown signal is a linear combination of a \textit{known} number of complex exponentials corrupted by additive white Gaussian noise.  The first step in a \ac{MUSIC} approach to \ac{QNS} is to again compute an estimate of  autocovariance matrix $\hat{\bm{R}}$.
Since $\hat{\bm{R}}$ is symmetric, it admits the eigendecomposition $\bm{V}\bm{\Lambda}\bm{V}^\top$, where $\bm{V}$ is orthogonal (i.e., real-valued and unitary), and $\bm{\Lambda}$ is a diagonal matrix of nonnegative eigenvalues $\lambda_i$, with $\lambda_1\geq\lambda_2\geq\cdots\geq\lambda_J\geq0$. Under the assumption that $\eta$ consists of $p$ complex exponentials, \ac{MUSIC} first defines a ``pseudo-spectrum'' by
\begin{equation}\label{eq:MUSIC}
  \hat{S}_{MU}(\omega)=\frac{1}{\sum_{i=p+1}^J|\bm{e}^\dagger(\omega)\bm{V}_i|^2}\,,
\end{equation}
where $\bm{V}_i$ denotes the $i$th column of $\bm{V}$ and $\bm{e}(\omega)=(1, e^{i\omega}, e^{i2\omega}, \dots, e^{i(J-1)\omega})^\top$, a frequency vector. Next, the $p$ largest local maxima of $\hat{S}_{MU}$ (between $0$ and $\pi$) are determined numerically and set as the (normalized) frequencies of the $p$ complex exponentials assumed to compose the signal.

\subsection{Model Based Spectrum Estimation Using SchWARMA}
The \ac{SchWARMA} approach introduced in \cite{schultz2021schwarma} and experimentally verified in \cite{murphy2021universal} provides a link between classical \ac{ARMA} models and noisy quantum circuits. In the restricted setting of Eq.~\eqref{eq:depham}, the continuous-time dynamics are replaced by a noisy discrete-time circuit with propagator
\begin{equation}
U(KT)=\prod_{k=1}^K Z(y[k])X\left(\int_{(k-1)T}^{kT}\,dt\Omega(t)\right),
\end{equation}
where $K$ is the number of gates in the noiseless circuit, $T$ is the gate time, $y[k]$ are the outputs of a classical ARMA model as in
Eq.~\eqref{eq:ARMA}, and $Z(\theta)$ and $X(\theta)$ are $Z$ and $X$ rotation
gates, respectively.  Assuming the resulting $X(\theta)$ have
$\theta\in\{0,\pi\}$ (as in the \ac{FTTPS}), the resulting filter function computation of Eq.~\eqref{eq:surv} is exact  in the instantaneous control limit.

Given the results $\{p_k\}$ of some dephasing \ac{QNS} experiment, we can attempt to either fit an \ac{ARMA} model
using the methods described above, and thus produce a \ac{SchWARMA} model, or to fit the 
\ac{SchWARMA} model directly to the data $p_k$ without first performing an
\ac{NNLS} \ac{PSD} estimate.  To achieve the latter, we note that the survival
probabilities $p_k$ are differentiable with respect to the \ac{SchWARMA} model
parameters $a_i$ and $b_j$, as the power spectrum in Eq.~\eqref{eq:ARMAPSD} is
differentiable with respect to the model parameters and Eq.~\eqref{eq:surv} is
differentiable with respect to the power spectrum.  Thus, the SchWARMA-based
prediction $\hat{p}_k$ of a given \ac{QNS} probe sequence using
Eqs.~\eqref{eq:ARMAPSD} and \eqref{eq:surv} are differentiable with respect to the SchWARMA model parameters. Given a differentiable loss function $L(\{p_k\},\{\hat{p}_k\})$ of the true and predicted survival probabilities (such as $\sum_k (p_k-\hat{p}_k)^2$), it is possible to use gradient descent to minimize this loss. To this end, we use TensorFlow's autodifferentiation \cite{tensorflow2015-whitepaper} and its implementation of the ADAM optimizer \cite{kingma2014adam} to minimize the mean squared error of the predictions.

\section{Spectrum Estimation}\label{sec:se}
In this section, we investigate the \ac{SchWARMA} approach to spectrum estimation in a variety of scenarios, using both simulated and experimental data. We chose to focus only on the \ac{SchWARMA} approach as we generally found it to be more accurate from random initial conditions than the adaptations using the \ac{NNLS}-derived autocovariance.  Furthermore, the differentiability of the \ac{SchWARMA} approach ensures that it can improve (in a mean-square error sense) initial models estimated using those approaches.  To this end, we use the cepstrum-based estimates as our initial conditions for the \ac{SchWARMA} fits and this generally outperforms random initial conditions, especially for complex models. Additionally, the \ac{SchWARMA} approach is less reliant on the conditioning of the \ac{NNLS}-based regression problem and could potentially be used in conjunction with probe sequences that are not otherwise suited to effective \ac{QNS}, although we leave this investigation to future work.  However, the \ac{SchWARMA} approach does add computational and time overhead to computing an estimate, and it is possible that there are real-time \ac{QNS} problems that could benefit from the other adaptations. In these cases, we recommend the cepstrum-based approach as it fully subsumes the Yule-Walker approach for purely autoregressive models and scales much better for moving average models than the pure-optimization approach described above.

\subsection{White Noise}
\begin{figure}[ht!]

  \centering

  \includegraphics[width=\columnwidth]{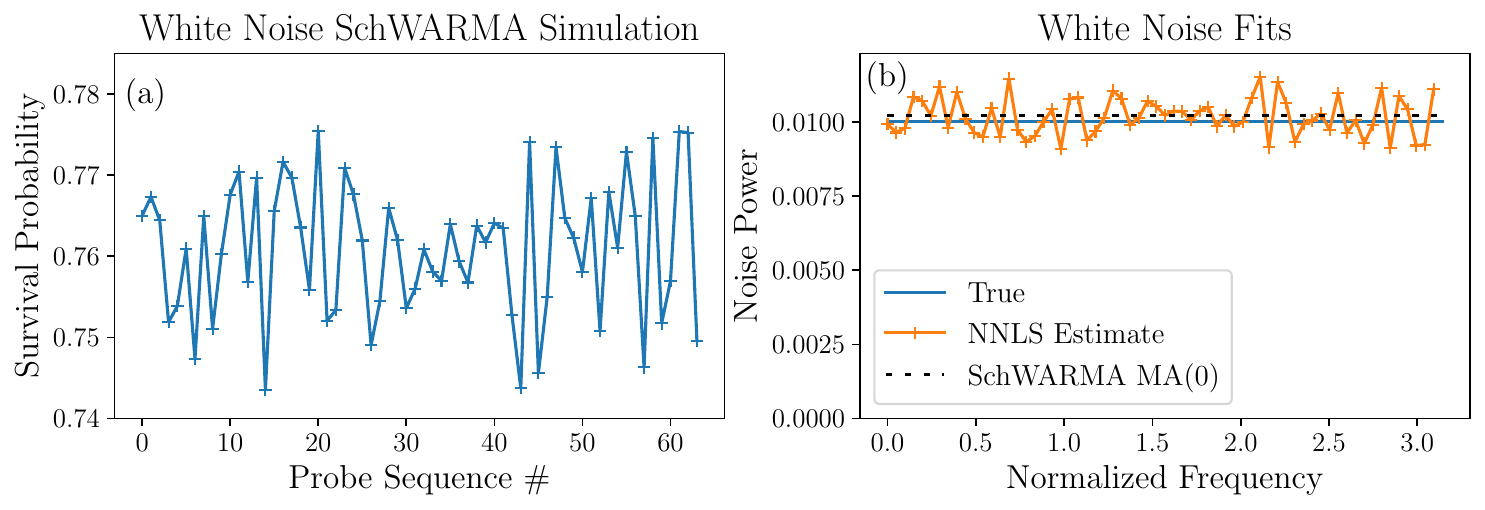}

  \caption{Simulated QNS experiment of white noise. (a) Simulated \ac{FTTPS} survival probabilities using 1000 \ac{SchWARMA} trajectories; note the variability due to the binomial measurement distribution. (b) True and estimated noise spectra.}
  \label{fig:white_noise}
\end{figure}

A simple example that highlights the difference between standard \ac{NNLS}-based spectrum estimation and model based approaches occurs when the dephasing noise spectrum comprises only white noise. In this case, the overlap integral in Eq.~\eqref{eq:surv} will be
constant for each sequence in the \ac{FTTPS}, since their total time is the
same. In the infinite measurement limit, we would therefore expect the survival
probabilities to be the same; however, when the measurements are finite, there
will be variance in the empirical survival probabilities due to the underlying
binomial distribution, as in Fig.~\ref{fig:white_noise}a.  When these survival
probabilities are inverted using \ac{NNLS} we see similar variation in the
estimated power in each band, as shown in Fig.~\ref{fig:white_noise}b.

In contrast, a model-based approach using a \ac{SchWARMA} model is able to more accurately estimate the true spectrum.  Since white noise is equivalent to an $MA(0)$ model (and indeed these simulations were performed using an $MA(0)$ \ac{SchWARMA} model), we should in principle be able to accurately estimate the spectrum using a model of the same form. When we fit a \ac{SchWARMA} model directly to the simulated \ac{QNS} experiment, we see that the power spectrum is flat and also quite accurate (see Fig.~\ref{fig:white_noise}b), here estimating the noise power within $\approx 0.7\%$. This simple example illustrates a potential benefit of model-based spectrum estimation. As there are now fewer model parameters than probe sequences, the experimental data is effectively ``pooled,'' reducing estimation error. To further demonstrate how this works, when we simply take the average survival probability across all probe sequences, we find that the estimated power spectrum from that average is comparable (sometimes better, sometimes worse) to the \ac{SchWARMA} estimate.

We note that in this case, the \ac{SchWARMA} fit is somewhat unnecessary, since it is fairly obvious that the spectrum in Fig.~\ref{fig:white_noise} is quite flat, and one would be inclined to fit a horizontal line to the resulting \ac{NNLS} estimates to determine the white noise power. A similar approach is often used in the fitting of $1/f^{\alpha}$ spectra to QNS experiments, whether the underlying estimates come from \ac{FTTPS} sequences \cite{murphy2021universal} or other probe approaches~\cite{alvarez2011:qns, yan2013:fnoise, pazsilva2017:qns, norris2018optimally}.

\subsection{Time-Correlated Noise}

\begin{figure}[ht!]

  \centering

  \includegraphics[width=\columnwidth]{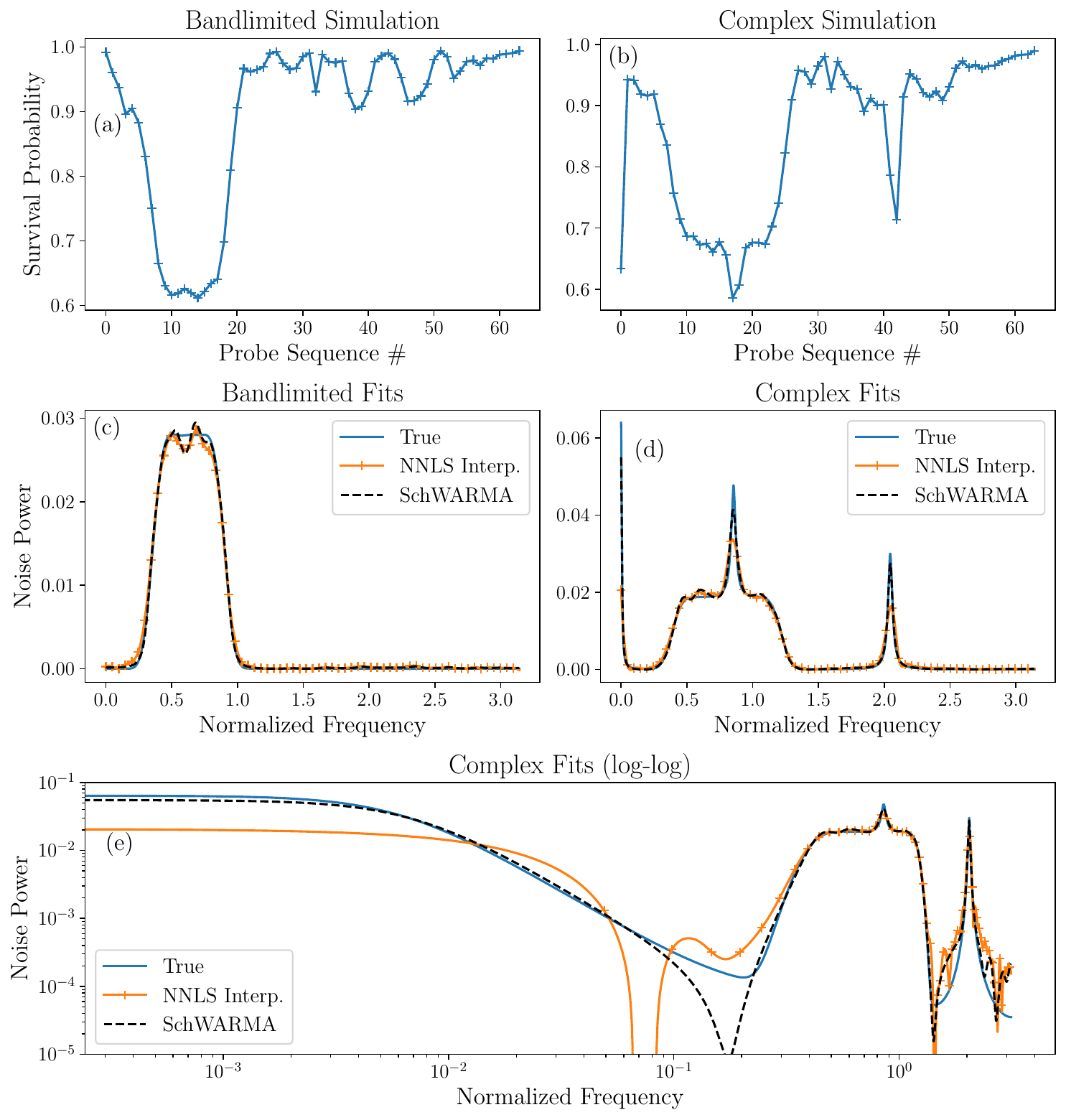}

  \caption{\ac{SchWARMA} simulations and spectral reconstructions for correlated noise spectra. (a) Simulated \ac{QNS} experiments of the bandlimited noise spectrum. (b) Simulated \ac{QNS} experiments of the complex noise spectrum formed by $1/f^2$ with additional spectral features. (c) Bandlimited noise spectrum and estimated spectra using the \ac{NNLS} and \ac{SchWARMA} approaches. (d) Complex noise spectrum and estimated spectra using the \ac{NNLS} and \ac{SchWARMA} approaches. (e) Same spectra as in (d), plotted on a log-log axis to better show the $1/f^2$ portion of the spectrum.}
  \label{fig:spectral_plots}
\end{figure}

To illustrate model-based spectrum estimation on more complex noise models, in this section we consider two different correlated noise spectra.  These are 1) a bandlimited noise spectrum, modeling the noise injection experiments of \cite{murphy2021universal} and 2) a complex noise spectrum generated by the sum of $1/f^2$ noise, a wide bandlimited region, and two narrow spectral features.  Simulated \ac{FTTPS} \ac{QNS} results for these two noise models are shown in Fig.~\ref{fig:spectral_plots}a and b, respectively. Noise reconstructions for the bandlimited case are shown in Fig.~\ref{fig:spectral_plots}c, and two different views for the complex spectrum are shown in Fig.~\ref{fig:spectral_plots}d and e.

For the \ac{NNLS} noise estimates, we performed cubic spline interpolation to produce spectrum estimates that are continuous in frequency, as opposed to the discrete point estimates naturally produced by the least-squares inversion process, in order to provide a more fair comparison between approaches.  Additionally, we found that cubic spline interpolation produced better results than lower-order spline interpolation and Fourier ($\sinc$)-based interpolation. We compare the accuracy of these two approaches later in this section, but for now we note that the spectral leakage of the filter functions in the \ac{NNLS} appear to result in slightly narrower estimates of the passband region at the top along with slightly wider estimates at the bottom, as shown in Fig.~\ref{fig:spectral_plots}c and e. Similarly, the \ac{NNLS} method appears to underestimate the peak of the spectral features in Fig.~\ref{fig:spectral_plots}d to a greater degree than the \ac{SchWARMA} approach.

As suggested in the previous section, a powerful property of \ac{ARMA} models that \ac{SchWARMA} models inherit is the ability to perform model selection, meaning the ability to select $p$ and $q$ in a way that balances error with overfitting \cite{stoica2004model}. Here, we will consider two procedures for model selection, \ac{AIC} and \ac{BIC}. When applied to (linear) regression problems, these approaches penalize some function of the log-liklihood in terms of the \ac{MSE} $1/K\sum_k(p_k-\hat{p}_k)^2$ of the fit residuals with a term that is dependent on the overall number of parameters (here $p+q+1$). Specifically, for this \ac{QNS} application, the \ac{AIC} of a \ac{SchWARMA} model is $K\log(MSE)-2(p+q+1)$, and the corresponding \ac{BIC} is $K\log(MSE)-(p+q+1)\log(K)$. Thus, the only difference between these two criteria is the way in which they penalize the number of parameters in the model.

\begin{figure}[ht!]

  \centering

  \includegraphics[width=\columnwidth]{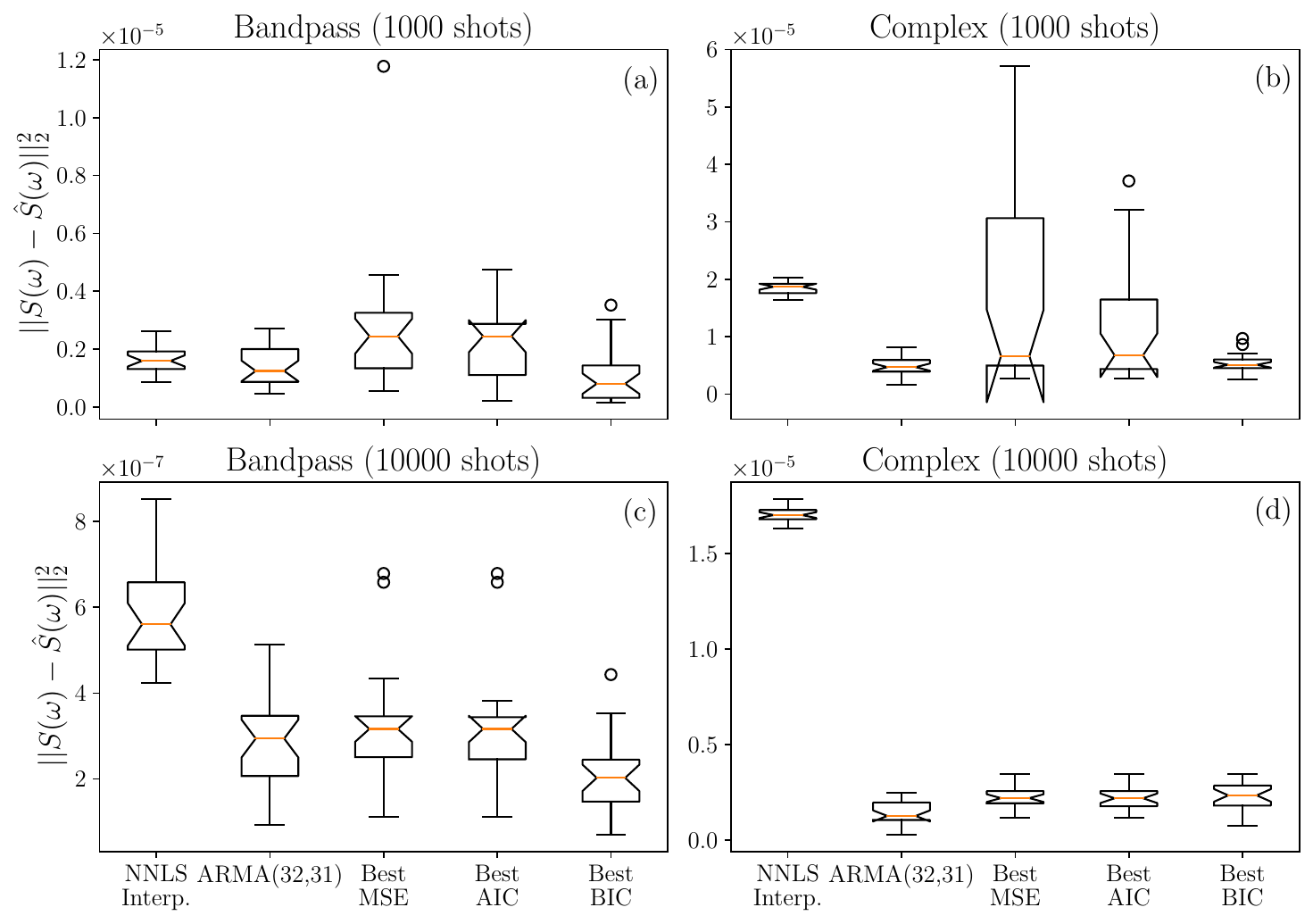}

  \caption{Boxplots (25 independent simulations) of spectrum reconstruction error for interpolated \ac{NNLS}, \ac{SchWARMA} fit $ARMA(31,32)$ models, and model-selected \ac{SchWARMA} models using minium mean squared error, Akaike information criterion, and Bayesian information criterion. (a) Bandpass noise, 1000 shots per probe sequence. (b) Complex spectrum, 1000 shots per probe sequence. (c) Bandpass noise, 10000 shots per probe sequence. (d) Complex spectrum, 10000 shots per probe sequence. Notches on the boxplot indicate 95\% confidence in differences between medians if they do not overlap.  Additionally, Wilcoxon signed-rank test indicates statistically significant differences between the \ac{NNLS}-interpolated estimates and the \ac{SchWARMA}-fit $ARMA(32,31)$ and best BIC selected models for each case.}
  \label{fig:mbqns_boxplots}
\end{figure}

To demonstrate and assess model selection while also validating \ac{SchWARMA} model-based \ac{QNS}, we performed 25 independent \ac{QNS} simulations of \ac{FTTPS} probe sequences under \ac{SchWARMA}-driven dephasing noise for each of the two noise classes. We considered both 1000 and 10000 measurement shots per probe sequence to asses the influence of binomially distributed measurement error on the results. We computed the \ac{NNLS} estimate for each \ac{QNS} simulation, then interpolated them to get a continuous estimate to compare to the true power spectrum.  Similarly, we fit \ac{SchWARMA} models for each $(p,q)$ pair such that $p+q+1\leq 64$ (the number of probe sequences). For each $(p,q)$, we computed the \ac{MSE} between the \ac{SchWARMA} predictions and the simulated data, along with the \ac{AIC} and \ac{BIC}.  We then selected the model that produced the minimum \ac{MSE}, \ac{AIC}, and \ac{BIC} and computed their spectrum reconstruction error.  Box plots of the spectrum error distributions are shown in Fig.~\ref{fig:mbqns_boxplots}.

The results of this Monte Carlo analysis demonstrate that using minimum BIC as a model selection criterion outperforms the interpolated \ac{NNLS} reconstructions in a statistically significant manner (Wilcoxon signed-rank test, max $\alpha<0.0004$ across the four scenarios), in contrast to using the minimum MSE or AIC as a selection criteria, which performed poorly on some of the 1000 shot trials. This is consistent with the general theory of model selection, as we expect \ac{BIC} to more accurately estimate the number of parameters in a model \cite{stoica2004model}.

We also recognize that some potential use cases of \ac{SchWARMA} model-based \ac{QNS} may not be amenable to exhaustive search over the range of valid $(p,q)$ choices, if, for example, it is being used as part of a fast adaptive control loop. To this end, we also include the $ARMA(32,31)$ model as a candidate that should exhibit flexibility in spectral structure with its mix of terms while also having the maximum number of parameters (here 64) in order to produce good fits. We find that this choice also outperforms the interpolated \ac{NNLS} method in reconstructing the true spectrum (Wilcoxon signed-rank test, max $\alpha<0.05$), with comparable performance to minimizing the \ac{BIC}.

\subsection{Native and Injected Experimental Noise in a Superconducting Qubit}
\begin{figure}[ht!]

  \centering

  \includegraphics[width=\columnwidth]{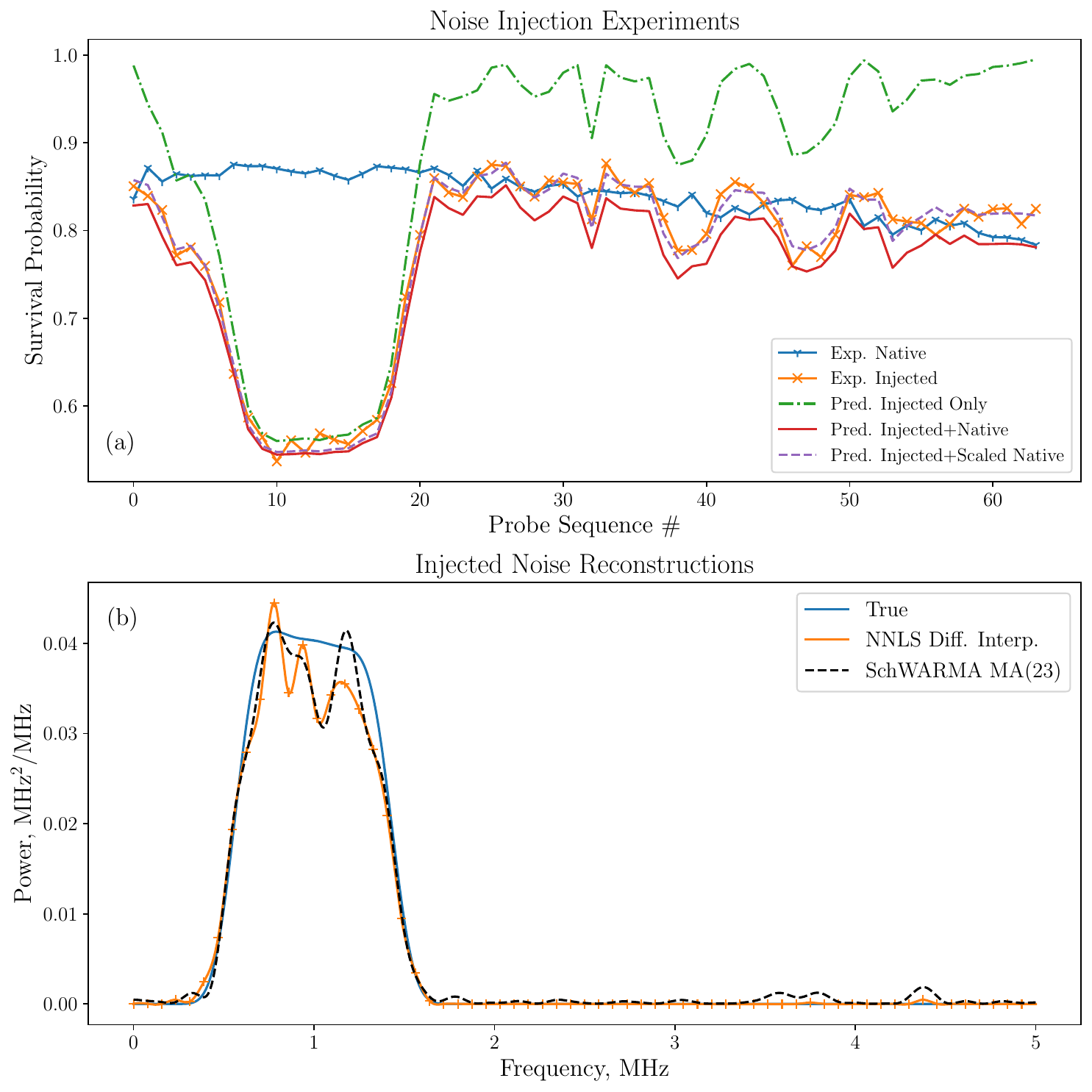}

  \caption{Analysis of noise injection experiment from Ref.~\cite{murphy2021universal}. (a) Survival probabilities of \ac{FTTPS} probe sequences with native and  bandlimited injected (SDR-based) noise from Ref.~\cite{murphy2021universal}, along with the \ac{SchWARMA}-based predictions that incorporate only the injected noise power, the sum of the injected noise power with a native noise fit, and the sum of the injected noise power with a scaled noise fit to account for native noise fluctuation observed in the data.  (b) Comparison of the true injected noise spectrum to an \ac{NNLS}-based method derived by subtracting and truncating the native noise \ac{NNLS} estimate, and the BIC-optimal \ac{SchWARMA} fit that includes a scaled native noise.}
  \label{fig:exp_data} 
\end{figure}

In Ref.~\cite{murphy2021universal}, a \ac{SchWARMA}-based scheme for experimental dephasing noise was introduced. In that work, it was demonstrated that \ac{SchWARMA} could accurately predict survival probabilities of \ac{QNS} experiments given a native noise model and a desired injected spectrum. In this section, we use the experimental data from Ref.~\cite{murphy2021universal} to demonstrate the applicability of the model-based spectrum estimation techniques to real experimental data. This effectively ``closes the loop'' on \ac{SchWARMA} as a noise simulation, injection, and estimation protocol. Fig.~\ref{fig:exp_data} shows bandlimited signal injected from Ref.~\cite{murphy2021universal}, along with various \ac{SchWARMA} predictions.

Clearly, the predictions of the injected noise by itself do not come particularly close to capturing the behavior of the combined native and injected signal, due to the strength of the native noise. To resolve this, we can first perform a \ac{SchWARMA} estimate of the native noise spectrum, $S_{nat}(\omega)$, then attempt to estimate only the injected noise spectrum, $S_{inj}(\omega)$, by replacing $S_\eta(\omega)$ in Eq.~\eqref{eq:surv} with the sum $S_{nat}(\omega)+S_{inj}(\omega)$. The native noise overlap term $-\int d\omega\,S_{nat}(\omega)F_{\Omega}(\omega)$ effectively behaves as a constant term in the \ac{SchWARMA} model optimization process, allowing for the fitting to proceed in essentially the same manner as before.
When we account for the native noise spectrum in this manner, for \textit{this specific} experimental data, it appears that there has been some drift between the \ac{FTTPS} experiments for native and injected \ac{QNS}. This is evident in the lower survival probability predictions for the combination of estimated native noise and true injected noise, especially outside the band of the injected signal, indicating an overestimation of the combined noise sources. 
To account for this variation and truly demonstrate the predictive power of \ac{SchWARMA}, we again use numerical optimization to find a native noise scale factor $\beta$ such that the sum of the known injected spectrum and a scaled estimate of the native noise $\beta S_{nat}(\omega)$ minimizes the MSE of the filter function predictions of the survival probability. Predicted survival probabilities of this combination of injected and scaled native noise are also shown in Fig.~\ref{fig:exp_data}a, and quite accurately predict the experimental results outside of the band of the injected signal.

With an accurate estimate of the native noise spectrum, we are now able to repeat the \ac{SchWARMA} fit using the scaled native noise. The actual injected power spectrum and the estimated injected power using the scaled native noise are shown in Fig.~\ref{fig:exp_data}b.
For reference, we also computed a truncated difference between the \ac{NNLS} estimates for the injected and native cases, setting all negative values to zero.
We note that it is (presumably) by pure chance that the native noise power decreased for the injection experiments, and this leads to a zero power outside of the injected band for the \ac{NNLS} reconstruction. If the situation were reversed, we would see spurious features in out-of-band portion of the estimated spectrum.

One interesting feature that we leave for future investigation is that the \ac{SchWARMA} model selection approach using \ac{BIC} drastically favors $AR(63)$ and $MA(64)$ models (i.e., the number of parameters are the same as the data size), for both the native and injected experimental data. This may be due to any number of reasons, including non-stationary fluctuations in the native noise and non-uniform accumulated gate error due to the differing number of $X$ gates in the \ac{FTTPS} sequences (as observed in Ref.~\cite{murphy2021universal}).  However, when we incorporate the scaled native spectrum information in the fit to the injected signal, we do find that an $MA(21)$ model minimized \ac{BIC} (which was used for the reconstruction in Fig.~\ref{fig:exp_data}).

\section{Superresolution}\label{sec:superres}
\begin{figure}[ht!]

  \centering

  \includegraphics[width=\columnwidth]{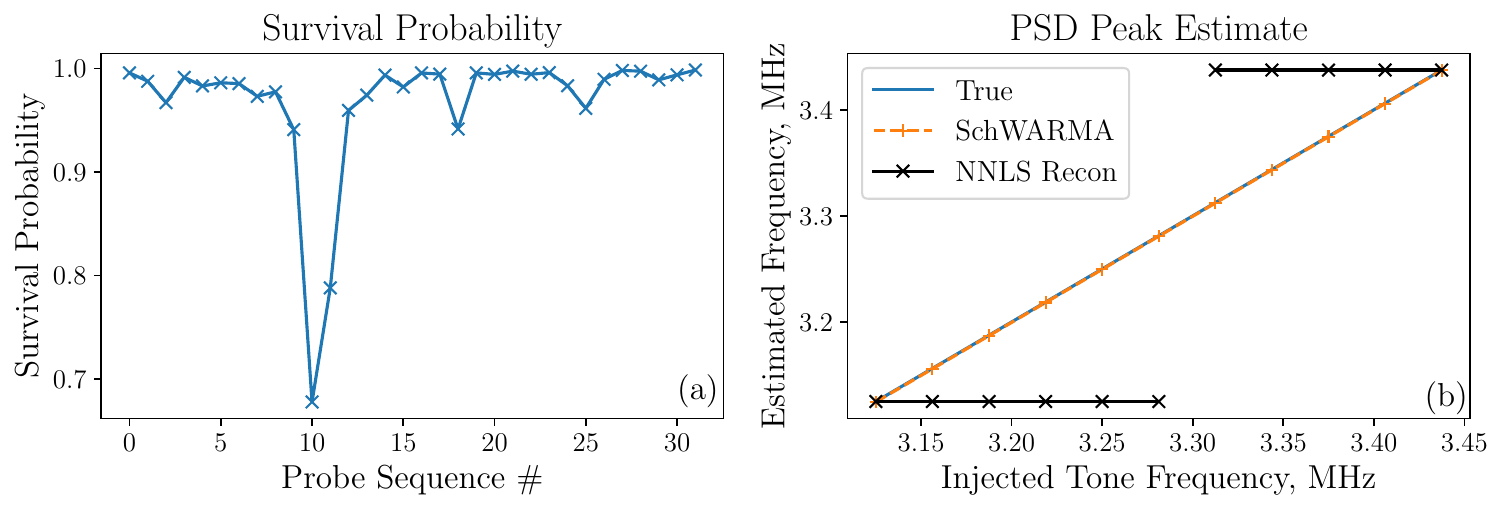}

  \caption{(a) Simulated \ac{FTTPS} \ac{QNS} experiment of an incoherent tone modeled by an $AR(2)$ model with a peak between two \ac{FTTPS} peak frequencies. (b) Frequency sweep of spectral peak and comparison of SchWARMA $AR(2)$ fit and \ac{NNLS} peak. }
  \label{fig:superres_fig1}
\end{figure}

As discussed above, \ac{NNLS}-based \ac{QNS} fundamentally estimates the power spectrum in a set of power bands, whereas the model-based approaches estimate the power (or presence of signal for \ac{MUSIC}) at individual frequencies.  Thus, these model-based \ac{QNS} approaches exhibit a phenomenon known as superresolution, where they are able to resolve spectral features at a finer scale than would be implied by the effective Nyquist rate.  Fig.~\ref{fig:superres_fig1}a shows \ac{SchWARMA}-simulated survival probabilities of \ac{FTTPS} dephasing spectrum with a peak between two \ac{NNLS} bins. Note that this noise spectrum produces a considerable drop in survival probability for two adjacent probe sequences.

\begin{figure}[ht!]

  \centering

  \includegraphics[width=\columnwidth]{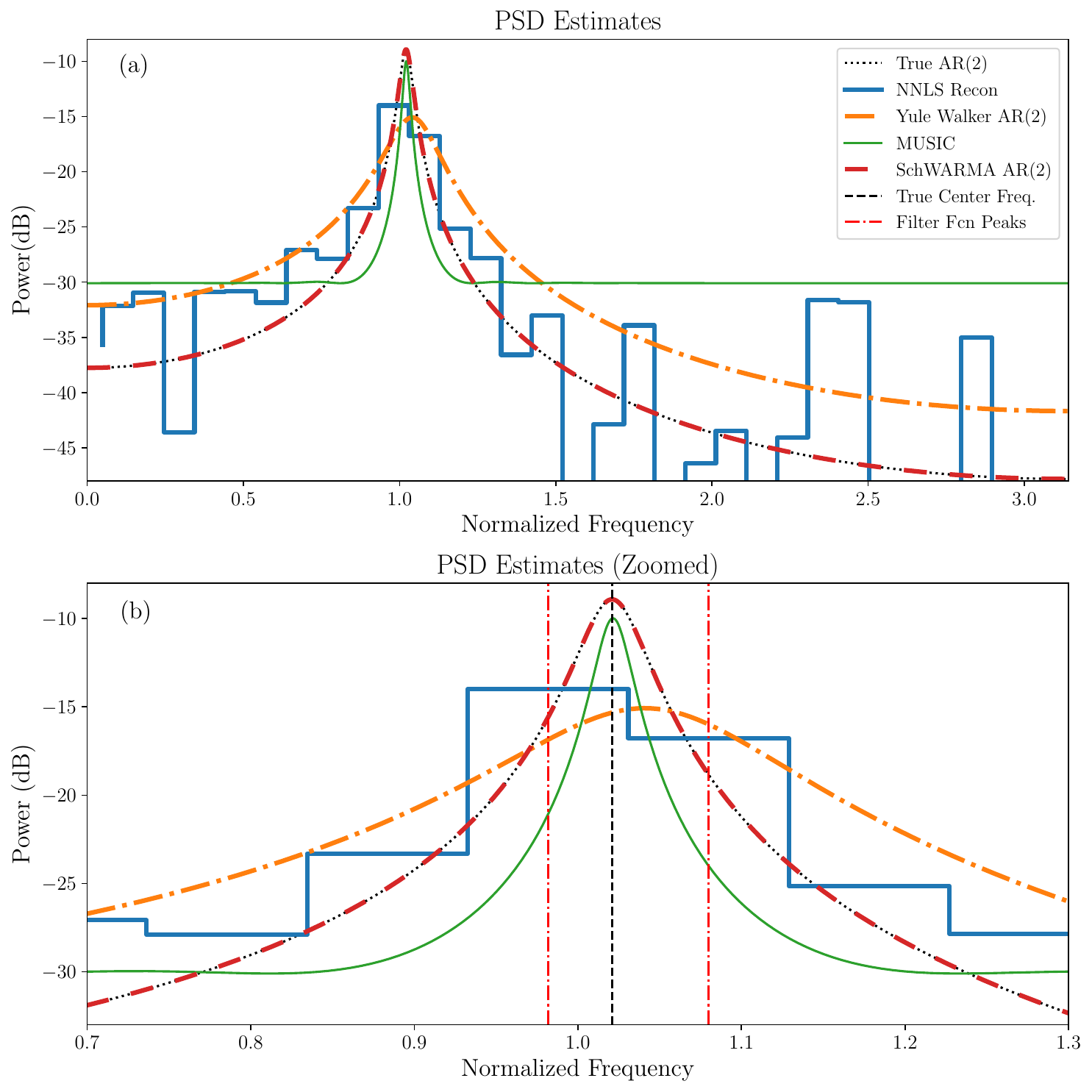}

  \caption{Spectrum estimation of a spectral peak from the simulation in Fig.~\ref{fig:superres_fig1}a. (a) Spectrum estimates over the full frequency range. (b) Zoomed view of spectrum estimates, with vertical lines indicating true center frequency ($--$) and peaks of the adjacent probe sequences ($-.$). }  \label{fig:superres_zoom}
\end{figure}

Fig.~\ref{fig:superres_zoom} shows spectrum estimates of this peaked spectrum using several different estimation approaches. As expected, the \ac{NNLS} reconstruction estimates considerable power in the two bands nearest the true center frequency. Using the \ac{NNLS} estimate to produce a Yule-Walker $AR(2)$ estimate results in a spectrum estimate that is not particularly accurate with respect to either the overall power or the peak center frequency. In contrast, the \ac{MUSIC} estimate using the same \ac{NNLS} reconstruction has an accurate estimate of the peak's center frequency. However, as noted above, the MUSIC estimate only produces a pseudo-spectrum from which the noise's constituent harmonic components can be estimated. From Fig.~\ref{fig:superres_zoom} we see that this pseudo-spectrum is not suitable for estimating the overall power spectrum.  Finally, a directly fit \ac{SchWARMA} AR(2) model produces a spectrum estimate that is extremely accurate to the true noise spectrum in terms of both the overall spectrum and the center frequency of the peak.

\begin{figure}[ht!]

  \centering

  \includegraphics[width=\columnwidth]{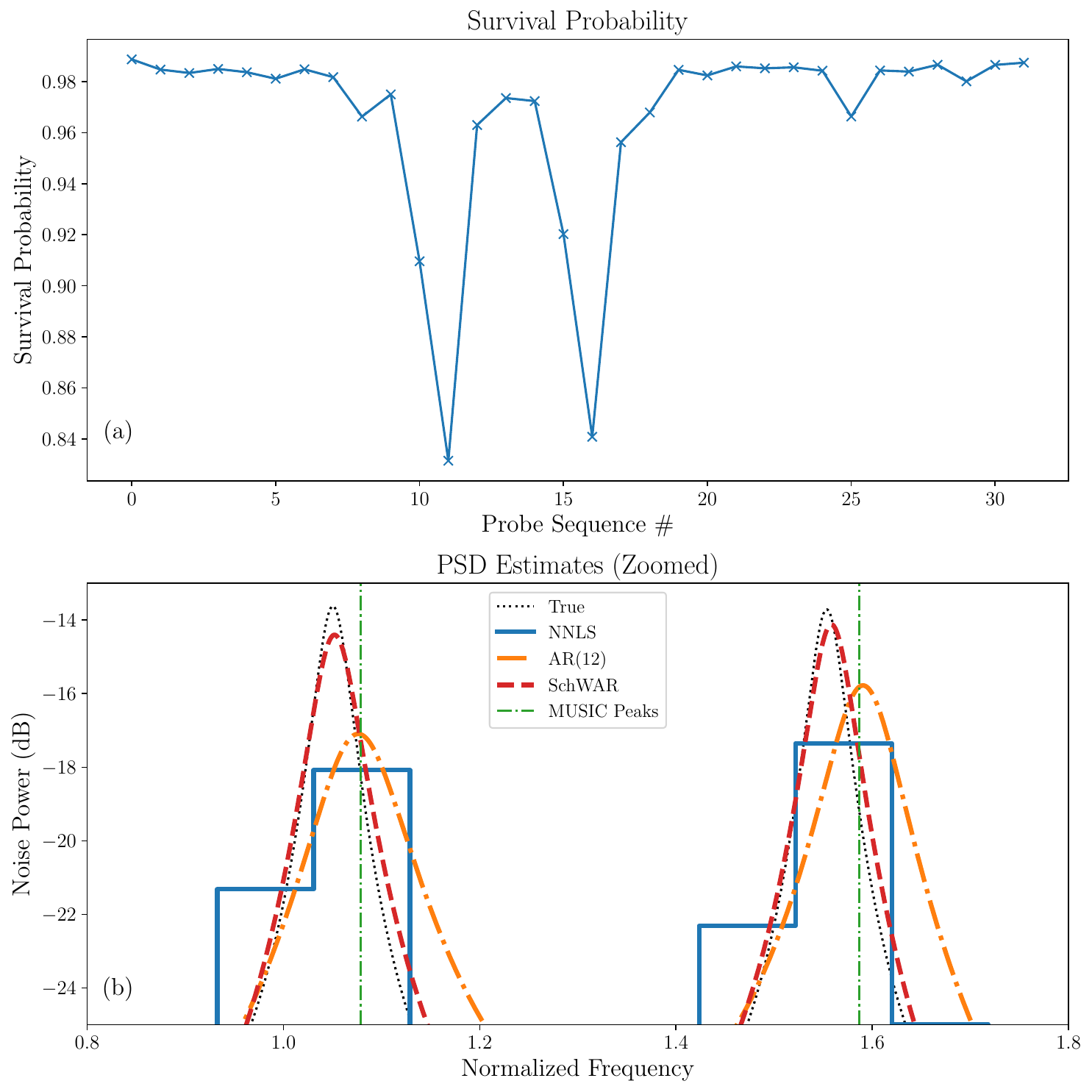}

  \caption{Superresolution of a multiple-peak spectrum. (a) \ac{FTTPS} survival probability simulation. (b) Comparison of different spectrum estimates to the survival probability data from (a).}
  \label{fig:two_tone}
\end{figure}

In the quantum context as with classical superresolution, these techniques function in the presence of multiple peaks (see Fig.~\ref{fig:two_tone}).
As before, the \ac{NNLS}-based estimates have spectral concentration in pairs of adjacent bands, and this appears to bias both the Yule-Walker and \ac{MUSIC} approaches.
Additionally, we observed that we had to increase the number of parameters in the Yule-Walker and SchWARMA AR fits in order to produce a spectral estimate that results in two peaks (here $AR(12)$ was used).  When the SchWARMA fits produce two peaks, the center frequencies of the peaks are quite accurate to the true center frequencies, although the overall power is not as accurate as the single peak case.

\subsection{Experimental Validation}

\begin{figure}[ht!]

  \centering

  \includegraphics[width=\columnwidth]{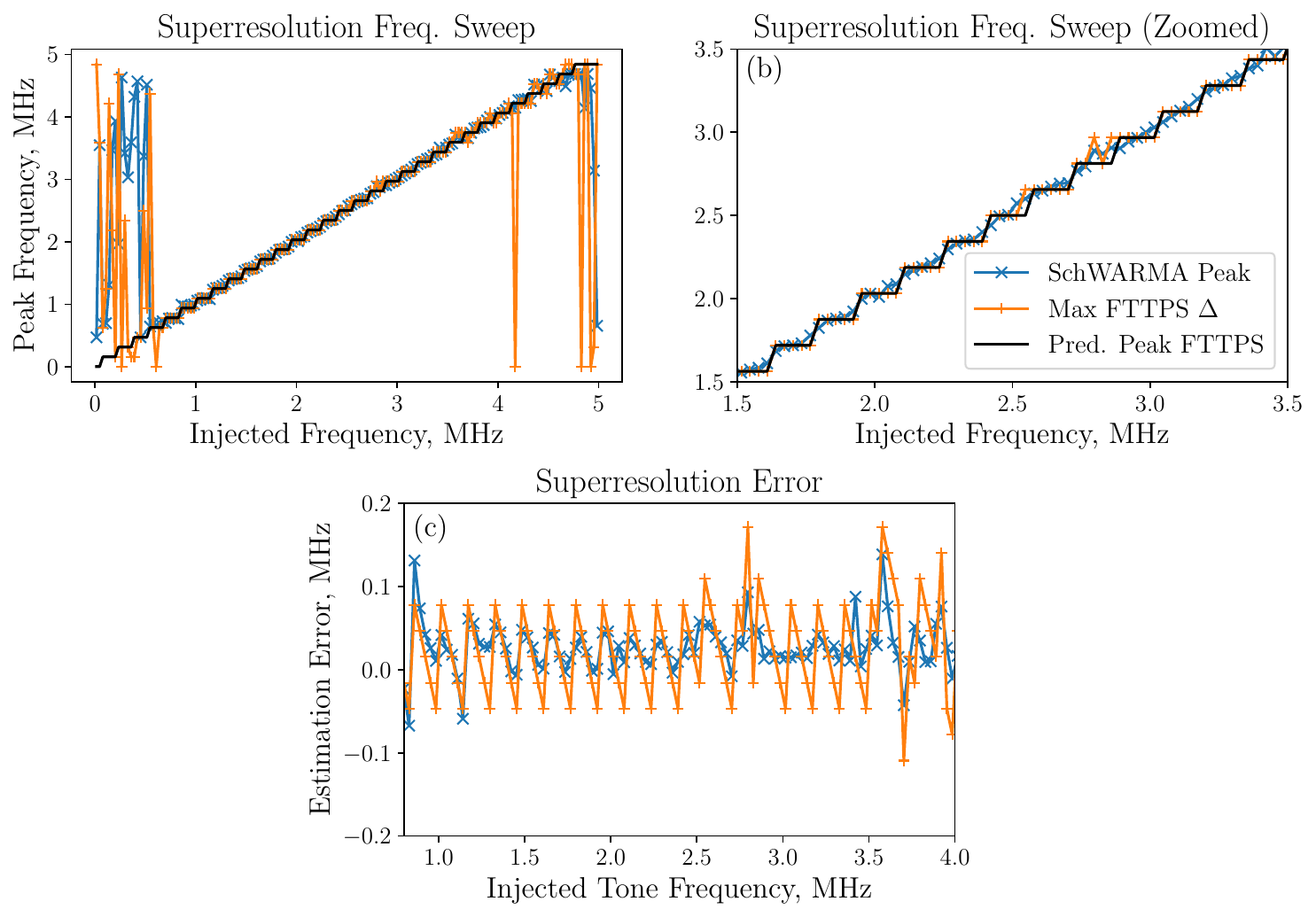}

  \caption{Experimental superresolution results of an injected spectral peak. (a) Full sweep of peak center frequency from 0-5~MHz in 31.25~kHz increments. (b) Zoomed in view. (c) Error of peak estimates of \ac{SchWARMA} and \ac{NNLS} reconstructions.}
  \label{fig:superres_exp_fig1}
\end{figure}

Fig.~\ref{fig:superres_exp_fig1} shows the results of an experimental demonstration of superresolution using \ac{SchWARMA} as compared to \ac{NNLS} reconstructions. For these experiments, we injected a spectral peak (itself generated by an $AR(2)$ model) with center frequency between 0-5~MHz in 31.25~kHz increments, one fifth the resolution of the \ac{FTTPS}-based \ac{NNLS} approach. This signal was injected into our superconducting transmon qubit system using the software-defined radio-based approach of Ref.~\cite{murphy2021universal}. To identify the center frequency of the injected peak, we first performed \ac{QNS} of the native noise using each approach, then identified either the bin (for \ac{NNLS}) or frequency (for \ac{SchWARMA}) that exhibited the biggest increase when the additional noise was injected.  For both spectrum estimation approaches, we see considerable distortion at the extremes of the frequency range. For low frequencies, we conjecture that this is due to fluctuations in the $1/f^\alpha$ component of the native noise spectrum, which is dominant at low frequencies. For the higher frequencies, this appears to be an artifact of the noise injection approach, whose output power (as measured by the survival probability of the qubit) drops off noticeably at higher frequencies. Despite these limitations, we see that both approaches track the predicted \ac{FTTPS} peak for a wide frequency range, and that the \ac{SchWARMA} approach is more accurate to the injected peak. We note that finite measurement effects as well as fluctuations in the native noise are likely the limiting factors in the overall accuracy of the \ac{SchWARMA} superresolution estimates.

\section{Conclusion}

In conclusion, we have adapted a number of different model-based approaches for classical spectrum estimation to the field of \ac{QNS}, demonstrating their efficacy in several scenarios using both simulation and experimental data. Specifically, we showed that \ac{SchWARMA}-based estimators produced more accurate power spectra than \ac{NNLS}-based estimators and that model-selection using \ac{BIC} can be used to reduce the impact of overfitting. Additionally, we showed that SchWARMA can resolve much finer spectral features than \ac{NNLS}-based estimation. Finally, we ``closed'' the \ac{SchWARMA} modeling and estimation loop, verifying that the \ac{SchWARMA} model is an effective model for noise simulation, injection, and estimation. Thus, the use of \ac{SchWARMA} as a spectral model offers several potential benefits over standard \ac{NNLS} approaches to \ac{QNS}, and we have only scratched the surface of potential adaptation of classical time-series analysis to the \ac{QNS} domain.  
%
%
We further note that one especially parsimonious use of the \ac{SchWARMA} approach to model-based \ac{QNS} is that the resultant SchWARMA model can immediately be used as a generative model of the noise in a quantum system, through which the impacts of the noise on various circuits can be explored as in \cite{schultz2021schwarma,schultz2022reducing} or used directly in model-based optimal control protocols \cite{trout2022provably}.

Several of the model-based \ac{QNS} approaches we considered are derived directly from a traditional non-parametric \ac{NNLS} estimate of the spectrum, as the limitations of quantum mechanics prevent the direct computation of the noise autocovariance. While these approaches still have utility despite the dependence on the \ac{NNLS} estimator, they are fundamentally limited by the statistical properties of that estimator, in particular the conditioning of the inversion matrix, which is ultimately determined by the probe sequences used.  The \ac{FTTPS} have excellent conditioning, but the number of sequences required (and their gate length) also scales linearly with the desired frequency resolution.  In contrast, direct estimation of an underlying \ac{SchWARMA} model is more flexible and could potentially produce useful, reduced order estimates despite ill-conditioning of the probe sequences used.  Furthermore, it should be possible to extract spectrum information from characterization experiments not generally designed for spectrum estimation, such as randomized benchmarking \cite{knill2008randomized} or gate set tomography \cite{Nielsen2021gatesettomography}. Since model-based spectrum estimation is ultimately about the number of model parameters to be estimated, the number of probe sequences required for accurate estimation should be on the order of the number of model parameters, and not the desired frequency resolution. Additionally, there is the obvious potential for compressive sensing-like algorithms that can use a few random sequences to extract meaningful spectral information.

With regard to the experimental noise-injection data, the primary focus was demonstrating that the \ac{SchWARMA} fitting approach can work with real experimental data, and indeed, we showed that we could find models that agreed with the survival probabilities up to numerical precision. However, it is likely that there are some non-dephasing noise sources that are also impacting survival probabilities, such as $T_1$ decay and gate-error, and a more complicated model would be required to discriminate between them. These unmodeled effects could also explain why the model-selection criteria tend to favor high parameter counts. Furthermore, the noise injection and native noise scenario is somewhat contrived in that we ``know'' the injected signal.  In a more realistic quantum sensing scenario focused on signal detection as in \cite{titum2021optimal}, we can envision a system that continuously performs \ac{QNS} and then tests the hypothesis that the native noise has changed or been rescaled against the possibility that an anomalous signal has been ``injected'' or is otherwise present in the system.  In this case, we feel that the model selection rules and other statistical tools that operate on candidate models will lead to effective solutions, as classical \ac{ARMA} models do in adaptive and statistical signal processing \cite{ingle2005statisical}.

\section*{Data availability}

Python implementations of the model-based \ac{QNS} routines are available at \href{https://github.com/mezze-team/mezze}{https://github.com/mezze-team/mezze}.

\section*{Acknowledgements}
 
KS and GQ acknowledge funding from ARO MURI grant W911NF-18-1-0218. GQ acknowledges funding from the U.S. Department of Energy (DOE), Office of Science, Office of Advanced Scientific Computing Research (ASCR), Accelerated Research in Quantum Computing program under Award Number DE-SC0020316.

\bibliography{references}

\end{document}